\documentclass[aps,pra,twocolumn,superscriptaddress,nofootinbib]{revtex4-2}
\usepackage{amsmath,amssymb,amsfonts,amsthm}
\usepackage{mathtools}
\usepackage{braket}
\usepackage{physics}
\usepackage{graphicx}
\usepackage{hyperref}
\usepackage{array}
\usepackage{tabularx}

\usepackage{bm}

\usepackage{amsthm}
\newcommand{\id}{\mathbb{I}}
\hypersetup{
colorlinks=true,
linkcolor=blue,
citecolor=blue,
urlcolor=blue
}
\usepackage{amssymb,amsfonts,amsmath,amstext,graphicx,hyperref}
\usepackage{xr-hyper}

\usepackage{tikz}
\usepackage{braket}
\usepackage{graphicx}

\usepackage[normalem]{ulem}
\usepackage{tabularx}
\def\bra#1{\langle{#1}|}
\def\ket#1{|{#1}\rangle}

\usepackage{braket}
\usepackage{mathtools}
\hypersetup{
	colorlinks=true,
	linkcolor=blue,
	filecolor=magenta,
        citecolor=blue,
	urlcolor=brown,
}
\usepackage{xcolor}

\begin{document}

\title{Quantum Illumination with Symmetry-Constrained Random Unitaries}
\author{Sooryansh Asthana}
\affiliation{Department of Physics, Indian Institute of Technology Bombay, Powai, Mumbai 400076, India}
\author{Sai Vinjanampathy}
\email{sai@phy.iitb.ac.in}
\affiliation{Department of Physics, Indian Institute of Technology Bombay, Powai, Mumbai 400076, India}
\affiliation{Centre of Excellence in Quantum Information, Computation, Science and Technology, Indian Institute of Technology Bombay, Powai, Mumbai 400076, India}
\date{\today}

\begin{abstract}
Quantum illumination provides a quantum advantage in detecting weakly reflecting objects embedded in a noisy environment, even when environmental noise destroys most of the initial entanglement. We investigate this advantage using Haar-random probe states constrained to symmetry-resolved subspaces. Employing tools from quantum channel discrimination and asymptotic hypothesis testing, we derive the discrimination exponents associated with Haar-random probe ensembles and identify the role of symmetry in determining their performance. We show that typical states drawn from fixed-charge sectors achieve the same asymptotic quantum-illumination advantage as maximally entangled probes. In particular, we show that the effective thermal-noise suppression and the corresponding Chernoff exponent are governed by the dimension of the accessible symmetry sector. Our results reveal that the operational resource underlying quantum illumination can be generalized from fine-tuned structure of a specific probe state to the existence of a large symmetry-protected correlation subspace. These findings establish a direct connection between quantum illumination, symmetry-resolved typicality, and quantum channel discrimination, and demonstrate that near-optimal quantum hypothesis testing resources can emerge naturally from generic many-body quantum states constrained by conservation laws.

\end{abstract}

\maketitle

\section{Introduction}
\label{sec:Introduction}
Quantum illumination (QI) is a quantum-assisted hypothesis testing protocol that retains a provable advantage over classical strategies in the presence of strong loss and environmental noise \cite{lloyd2008enhanced, shapiro2009quantum, tan2008quantum, Sanz17, fan2018quantum, lee2021quantum, karsa2022quantum}. Originally introduced via a simplified model \cite{lloyd2008enhanced} and subsequently generalized to Gaussian-state settings \cite{tan2008quantum}, QI addresses the task of detecting a weakly reflecting object embedded in a bright thermal background. Remarkably, the quantum advantage persists even though the signal--idler entanglement used during state preparation is almost completely destroyed by interaction with the environment. 

From an information-theoretic perspective, QI is naturally formulated as a problem of quantum channel discrimination  \cite{lloyd2008enhanced,subramanian2024shallow}. The presence or absence of a target corresponds to two competing quantum channels acting on a probe system, and the hypothesis testing task reduces to distinguishing these channels with minimum error probability. In the standard protocol \cite{lloyd2008enhanced}, entangled signal--idler probes together with joint measurements produce a suppression of the effective thermal noise, leading to an enhanced discrimination exponent and a corresponding reduction in detection error probability. The channel-discrimination viewpoint therefore provides a unifying framework for understanding both the origin and the limits of the quantum advantage.

A natural question is whether this advantage relies on carefully engineered entangled states or whether it can also emerge from more generic quantum resources. Haar-random states constitute a particularly appealing candidate in this regard.  Beyond their fundamental role in quantum information theory, Haar-random states have recently emerged as a resource in shadow-tomography protocols \cite{aaronson2018shadow, huang2020predicting} and learning-based approaches to quantum-state characterization \cite{chia2025quantum}. Typical Haar-random states possess near-maximal entanglement and frequently arise as effective descriptions of chaotic many-body quantum systems. One might therefore expect them to be highly effective probes for channel discrimination and sensing tasks. Haar-random states also exhibit strong concentration of measure and are locally close to maximally mixed states \cite{mele2024introduction}. In this work we show that Haar-random states can nevertheless achieve the full QI advantage (enjoyed by entangled states) when the randomness is restricted to an appropriate symmetry-resolved subspace. Specifically, we consider pure Haar-random states drawn from fixed-charge sectors associated with a conserved charge \cite{hearth2025unitary}. These \textit{random} states remain highly entangled while preserving the symmetry structure imposed by the conservation law. We demonstrate that such symmetry-constrained Haar-random probes reproduce the characteristic noise suppression of QI. More generally, the discrimination performance depends not on the fine-tuned probe state but on the dimension and symmetry of the subspace in which the relevant channel information is encoded. This reveals a form of universality in QI: the achievable advantage is determined primarily by coarse structural features of the accessible Hilbert space rather than by the fine-tuned probe state itself.

Beyond QI itself, the analysis developed here naturally extends to a broader class of channel-discrimination problems. We first formulate Haar-random probe ensembles within a classical--quantum framework in which the random realization is retained and subsequently decoded. This permits an exact  independent and identically distributed (i.i.d.) description of the resulting discrimination protocol and leads to a quantum-Chernoff characterization of the asymptotic error exponent. We then analyze the structure of this exponent for both unrestricted and symmetry-constrained Haar ensembles. In particular, we identify how the accessible Hilbert-space dimension and symmetry-resolved spectral properties determine the achievable discrimination rate.
 {In the special case of unitary channel discrimination, the resulting exponents acquire a direct connection to Haar averages, spectral form factors, and quantum-chaotic statistics which we elaborate on. QI subsequently emerges as an operational application of this more general framework.}

The results also suggest a broader perspective on quantum channel discrimination. The achievable asymptotic discrimination rate is governed not only by the properties of individual probe states but also by the structure and dimension of the subspace from which the probes are drawn. From this viewpoint, the relevant resource is the effective correlation subspace that remains distinguishable under the competing channel hypotheses. We prove that restricting probe states to symmetry-resolved sectors modifies this accessible space and consequently alters the associated discrimination exponent.   Our findings establish a direct connection between channel discrimination and contemporary developments in quantum many-body physics. Charge-conserving Haar-random states arise naturally as effective descriptions of eigenstates and long-time dynamics of chaotic systems with global conservation laws \cite{hearth2025unitary}. Consequently, the probe states considered here are not merely abstract mathematical constructions but are representative of states generated by generic ergodic dynamics. This places QI within a broader framework linking quantum sensing \cite{oszmaniec2016random, fiderer2018quantum, kobrin2024universal, asthana2025projected, asthana2026efficient}, symmetry-resolved scrambling \cite{hearth2025unitary}, quantum chaos, and complexity  \cite{cotler2017chaos}. In particular, our results suggest that near-optimal discrimination resources can emerge naturally from chaotic dynamics constrained by conservation laws, without the need for finely engineered entangled states.

The paper is organized as follows. In Sec.~\ref{sec:Quantum state and channel discrimination} we review quantum state discrimination, channel discrimination, and QI from the hypothesis-testing perspective. In Sec.~\ref{sec:haar_random_qi} we formulate channel discrimination with Haar-random signal–idler probes, establish the corresponding Chernoff asymptotics, and analyze the resulting discrimination exponents for both unrestricted and symmetry-constrained Haar ensembles. In Sec.~\ref{sec:symmetry_haar_chernoff} we investigate symmetry-constrained Haar-random probes in greater detail, deriving symmetry-resolved Chernoff exponents and establishing their connection to spectral form factors. We then specialize these results to the QI setting in Sec.~\ref{sec:QI_symmetry_constrained}, where we show how symmetry-resolved Haar-random states reproduce the characteristic noise suppression and discrimination advantage of conventional QI. In Sec.~\ref{sec:Discussion}, we discuss physical realizations based on number-conserving dynamics, relate our framework to randomness-assisted illumination protocols, and interpret the discrimination advantage in terms of symmetry-resolved correlation subspaces. Finally, in Sec.~\ref{sec:Conclusion and outlook} we summarize the implications of our results and discuss future directions.

\section{Quantum state and channel discrimination}
\label{sec:Quantum state and channel discrimination}
QI is fundamentally a problem of distinguishing between two competing physical hypotheses \cite{lloyd2008enhanced}. In this section, we briefly review the general framework of quantum state and quantum channel discrimination, which provides the theoretical foundation and summarizes notations for the analysis that follows.

\subsection{Quantum state and channel discrimination}
\label{subsec:Quantum state discrimination}
Quantum state discrimination concerns the task of identifying an unknown quantum state drawn from a known ensemble \cite{Audenaert07,nussbaum2009chernoff,Bavaresco21}. In the simplest binary setting, the system is prepared in either state $\rho_0$ or $\rho_1$ with prior probabilities $p_0$ and $p_1=1-p_0$, respectively. The objective is to perform a measurement that minimizes the probability of assigning the wrong label to the state \cite{chefles1998quantum,bergou2010discrimination}. A general binary quantum measurement is described by a positive-operator-valued measure (POVM) $\{M_0,M_1\}$ satisfying $M_0+M_1=\id $ and $M_i\ge 0$. The outcome associated with $M_0$ is interpreted as the decision that the state is $\rho_0$, while the outcome associated with $M_1$ corresponds to the decision that the state is $\rho_1$. Consequently, two types of errors can occur. If the true state is $\rho_0$, an error is made when the outcome corresponding to $M_1$ is obtained, which occurs with probability $\Tr(\rho_0M_1)$. Conversely, if the true state is $\rho_1$, an error occurs when the outcome corresponding to $M_0$ is obtained, with probability $\Tr(\rho_1M_0)$. These are the analogs of type-I and type-II errors in classical binary hypothesis testing \cite{neyman1928use, neyman1928use_2}. Averaging over the prior probabilities, the total probability of error is
$P_e=p_0\Tr(\rho_0M_1)+p_1\Tr(\rho_1M_0)$.
Optimizing this quantity over all POVMs yields the minimum achievable error probability. For a single copy of the unknown state, the optimal error probability is given by the Helstrom bound \cite{holevo1998coding,helstrom1969quantum},
$P_e^{\mathrm{opt}}=\frac{1}{2}\bigl(1-\|p_0\rho_0-p_1\rho_1\|_1\bigr)$,
where $\|A\|_1=\Tr\sqrt{A^\dagger A}$ denotes the trace norm. The optimal measurement is obtained by projecting onto the positive  eigenspaces of the Helstrom operator $p_0\rho_0-p_1\rho_1$.

When multiple independent copies of the state are available, i.e., $\rho_0^{\otimes n}$ {or}  $\rho_1^{\otimes n},$
the optimal error probability decreases exponentially with the number of copies,
$P_e^{(n)} \sim \exp({-n \xi}),$
where $\xi = -\min_{0 \le s \le 1} \log \Tr \left( \rho_0^s \rho_1^{1-s} \right)$ is the quantum Chernoff exponent \cite{Audenaert07,helstrom1969quantum}. The exponential suppression in error probability is natural: if we consider two quantum states whose overlap is $\varepsilon$, the $n$-copy tensor product states are more orthogonal as their overlap goes as $\varepsilon^n$. This increased orthogonalization provides a natural advantage in state (and hence channel) discrimination. We note that these results can furthermore be related to large-deviation theory \cite{audenaert2008asymptotic}. 

It is important to note that achieving the optimal exponent generally requires collective measurements acting jointly on all copies \cite{Audenaert07}. Such measurements can exploit quantum correlations across copies and outperform any strategy based on independent measurements. {It is also worth noting the counterpoint, namely that saturation of the quantum Chernoff exponent does not always require collective measurements. For example, in binary coherent-state discrimination, repeated independent applications of the Kennedy detector \cite{kennedy1973near} attain the optimal asymptotic error exponent, equal to the quantum Chernoff exponent, despite being a local shot-by-shot measurement strategy. This demonstrates that suitably chosen separable measurements can, in certain cases, achieve the asymptotically optimal discrimination rate.
} Beyond its role in collective measurement strategies, entanglement can also enhance state discrimination through the use of ancillary systems and entangled input states in channel-discrimination protocols, where the resulting correlations amplify differences between the output states and improve distinguishability \cite{bae2015quantum}.

Quantum channel discrimination generalizes state discrimination by replacing states with dynamical processes. In the binary setting, an unknown quantum channel is either $\Phi_0$ or $\Phi_1$, each acting on a system $A$ and occurring with prior probabilities $p_0$ and $p_1$. A general discrimination strategy consists of three steps: (a) Prepare an input state on a composite system consisting of the probe $A$ and a reference system $R$,  $\rho_{RA}$. (b) Apply the unknown channel to subsystem $A$. (c) Perform a joint measurement on the output state.
The resulting output states are
$\rho_i^{\mathrm{out}} = (\mathbb{I}_R \otimes \Phi_i)(\rho_{RA}),i=0,1,$
and the task of channel discrimination reduces to that of discriminating these two states.
Optimizing over all input states and measurements yields the minimum error probability $P_e^{\mathrm{opt}} =\left(1 - \| p_0 \Phi_0 - p_1 \Phi_1 \|_{\diamond} \right)/2,$ where $\|\cdot\|_{\diamond}$ denotes the diamond norm \cite{aharonov1998quantum}. This norm quantifies the maximum distinguishability of two channels, optimized over all possible input states, including those entangled with arbitrarily large reference systems.

The presence of the diamond norm highlights a key distinction between state and channel discrimination: entangled input states can strictly increase the distinguishability of channels. Entangled probe states can increase channel distinguishability beyond what is achievable with separable inputs \cite{Piani09}. This feature underlies many quantum advantages in communication and sensing.

If multiple uses of the unknown channel are available, one may consider either parallel (non-adaptive) or adaptive strategies. In the parallel setting, the channels act independently on different subsystems, $\Phi_0^{\otimes n}$ {or} $\Phi_1^{\otimes n}.$
The error probability again decays exponentially, $P_e^{(n)} \sim \exp({-n \xi_{\mathrm{ch}}}),$
where $\xi_{\mathrm{ch}}$ is the channel Chernoff exponent \cite{Audenaert07}.
While adaptive strategies can provide advantages in certain settings, the parallel scenario already captures the essential role of entangled probe states and collective measurements. In particular, entanglement can enhance the distinguishability of quantum channels by increasing the separation between the corresponding output states.

An important application of quantum channel discrimination is QI  \cite{lloyd2008enhanced}, which aims to detect a weakly reflecting target in a bright thermal background. The problem can be formulated as binary quantum hypothesis testing between two channels corresponding to target absence and presence, with the performance determined by the distinguishability of the resulting output states.
\subsection{Quantum Illumination as a Channel Discrimination Problem}
\label{subsec:quantum_illumination_channel_discrimination}

QI employs a bipartite signal--idler probe. The signal subsystem \(S\) is transmitted toward a target region, while the idler subsystem \(I\) is retained at the receiver. A joint measurement is subsequently performed on the returned signal and the retained idler to determine whether the target is absent or present. We briefly review the simplified single-photon model of Ref.~\cite{lloyd2008enhanced}.
Let $\rho_{SI}$ denote the signal--idler input state, $\rho$ the reduced signal state, $\eta\ll1$ the target reflectivity, \(b\) the average number of thermal photons per mode, and $d$ the number of available temporal, spectral, or spatial modes. In the low-noise regime, $\eta\ll1, b\ll1, bd\ll1,$
the thermal background state is well approximated by
\begin{equation}
\rho_{\mathrm{th}}
\simeq
(1-bd)\ket{\mathrm{vac}}\bra{\mathrm{vac}}
+
b\sum_{k=1}^{d}\ket{k}\bra{k},
\end{equation}
indicating that the background is dominated by the vacuum with a small probability of producing a single photon in one of the $d$ orthogonal modes.

The absence and presence of the target are represented by quantum channels acting on the signal subsystem
\begin{align}
\label{Lloyd Channel pair_1}
H_0:\quad
\Phi_0(\rho)
&=
\rho_{\mathrm{th}},
\\
H_1:\quad
\Phi_1(\rho)
&=
(1-\eta)\rho_{\mathrm{th}}
+
\eta \rho .
\label{Lloyd Channel pair}
\end{align}
Under $H_0$, the transmitted signal is completely lost and replaced by thermal noise. Under $H_1$, a fraction $\eta$ of the signal is reflected back to the receiver, while the remaining fraction is replaced by thermal noise.

The corresponding signal--idler output states are
\[
\rho_{SI}^{(i)}
=
(\Phi_i\otimes \id)(\rho_{SI}),
\qquad i\in{0,1},
\]
and the target-detection problem reduces to discriminating between $\rho_{SI}^{(0)}$ and $\rho_{SI}^{(1)}$. To quantify the advantage provided by entanglement, it is useful to first establish the performance of the optimal classical strategy.

\noindent{\it Classical illumination:} Consider first a classical strategy employing an unentangled single-photon probe $\ket{1}$. The receiver attempts to determine whether a detected photon originated from the reflected signal or from thermal noise. The output states become $\rho_0=\rho_{\mathrm{th}},
\rho_1
=
(1-\eta)\rho_{\mathrm{th}}
+
\eta \ket{1}\bra{1}.$ The corresponding click probabilities are $p(\mathrm{yes}|H_0)
= b,
p(\mathrm{yes}|H_1)
= b(1-\eta)+\eta.$  { The target increases the click probability by an amount
$\Delta p = p(\mathrm{yes}|H_1)-p(\mathrm{yes}|H_0)\simeq \eta$,
while the background noise level is set by the thermal-click probability $b$. Hence the effective classical signal-to-noise ratio (SNR) scales as
$\mathrm{SNR}_{\mathrm{cl}}\sim \eta/b$.} For $n$ independent transmissions, the optimal error probability obeys $P_e^{(n)}
\sim
\exp({-n\xi_{\mathrm{cl}}}),$
with Chernoff exponent $\xi_{\mathrm{cl}}
=
-\log
\left[
\min_{0\le s\le1}
\Tr
\left(
\rho_0^s
\rho_1^{1-s}
\right)
\right].$ In the low-noise limit, one finds $\xi_{\mathrm{cl}}
\simeq
\eta ,
(\eta>b),$ 
whereas in the noise-dominated regime $\eta<b$, $\xi_{\mathrm{cl}}
\simeq
{\eta^2}/{8b}.$

\noindent{\it Entangled QI:} A modified QI protocol has been proposed using an entangled signal--idler pair of Schmidt rank $d$, $\ket{\Psi}_{SI}
=
\sum_{k=1}^{d}
\ket{k}_S\ket{k}_I/{\sqrt d}$ \cite{lloyd2008enhanced}. Under $H_0$, the signal is completely replaced by thermal noise namely $\rho_{SI}^{(0)}
=
\rho_{\mathrm{th}}
\otimes{\mathbb{I}_I}/{d}.$ Under $H_1$, $\rho_{SI}^{(1)}
=
(1-\eta)\rho_{SI}^{(0)}
+
\eta
\ket{\Psi}\bra{\Psi}.$ The optimal receiver projects onto the maximally entangled state, $\Pi_{\mathrm{yes}}
=
\ket{\Psi}\bra{\Psi},
\Pi_{\mathrm{no}}
=
\id-\Pi_{\mathrm{yes}}.$ The corresponding detection probabilities become $p(\mathrm{yes}|H_0)
= {b}/{d},
p(\mathrm{yes}|H_1)=\eta+ (1-\eta){b}/{d}.$ { Thus the probability that a thermal photon is incorrectly identified as a returned signal photon is reduced from order $b$ to order $b/d$.} Equivalently,
$\mathrm{SNR}_{\mathrm{QI}}
\sim
{\eta d}/{b},$
corresponding to an enhancement by a factor of $d$ relative to classical illumination.  The associated Chernoff exponent becomes $\xi_{\mathrm{QI}}
\simeq
\eta,
 (\eta > b/d),$
while in the bad regime $\xi_{\mathrm{QI}}
\simeq
{\eta^2 d}/{(8b)}.$ Comparing with the classical result,
${\xi_{\mathrm{QI}}}/{\xi_{\mathrm{cl}}}
\simeq d$ in the bad regime, 
demonstrating the exponential advantage in the error probability  \cite{lloyd2008enhanced}.
An important insight of this analysis is that the advantage does not arise from entanglement surviving transmission through the noisy channel. In fact, under $H_0$ all signal--idler entanglement is completely destroyed. Rather, the advantage originates from the structure of the joint measurement performed at the receiver.
The maximally entangled state occupies a single direction within a $d$-dimensional correlation subspace, whereas thermal noise is approximately uniformly distributed throughout that subspace. Consequently, the overlap between noise and signal is suppressed by a factor of $1/d$, producing the enhancement in the Chernoff exponent.
Viewed from the perspective of channel discrimination, QI therefore exploits probe states that maximize the distinguishability between the channels $\Phi_0$ and $\Phi_1$. This interpretation suggests that the essential resource is not necessarily the specific spontaneous parametric down-conversion generated state employed in the original protocol \cite{kwiat1995new}, but rather the existence of highly structured correlations that permit efficient discrimination of the underlying channels \cite{brougham2023using}.
\section{Channel discrimination with Haar-random signal--idler probes}
\label{sec:haar_random_qi}

Having recast QI as an entanglement-assisted channel discrimination problem, it is natural to ask whether the specific maximally entangled probe employed in the original protocol is essential for the observed advantage. A particularly appealing alternative is provided by Haar-random states. Here we retain the idea of a \textit{signal and idler} though in practice these are just two subsystems, one sent to probe the unknown channel and another one retained, without any reference to any specific parametric down-conversion process. Typical pure states in high-dimensional Hilbert spaces possess nearly maximal bipartite entanglement with overwhelming probability quantified by, for instance, the Page value  \cite{Page1,Page2}. Furthermore according to this analysis sufficiently small subsystems of a Haar-random pure state are nearly maximally mixed \cite{hamma2012quantum}. Moreover, chaotic quantum dynamics are known to generate states that approximate Haar-random ensembles over a broad intermediate-time regime \cite{cotler2017chaos}. From this perspective, Haar-random states appear to contain an abundance of the resource traditionally associated with quantum-enhanced channel discrimination.

Motivated by these observations, we posit the following question: can a generic ensemble of Haar-random signal--idler states serve as an effective probe ensemble for QI and channel discrimination? If so, Haar-random states would provide a class of probes requiring little fine tuning. The purpose of this section is to formulate QI with Haar-random signal–idler probes and to identify the asymptotic channel-discrimination exponent governing the resulting hypothesis-testing problem.

\begin{figure}[hbt!]
    \centering
    \includegraphics[width=0.75\linewidth]{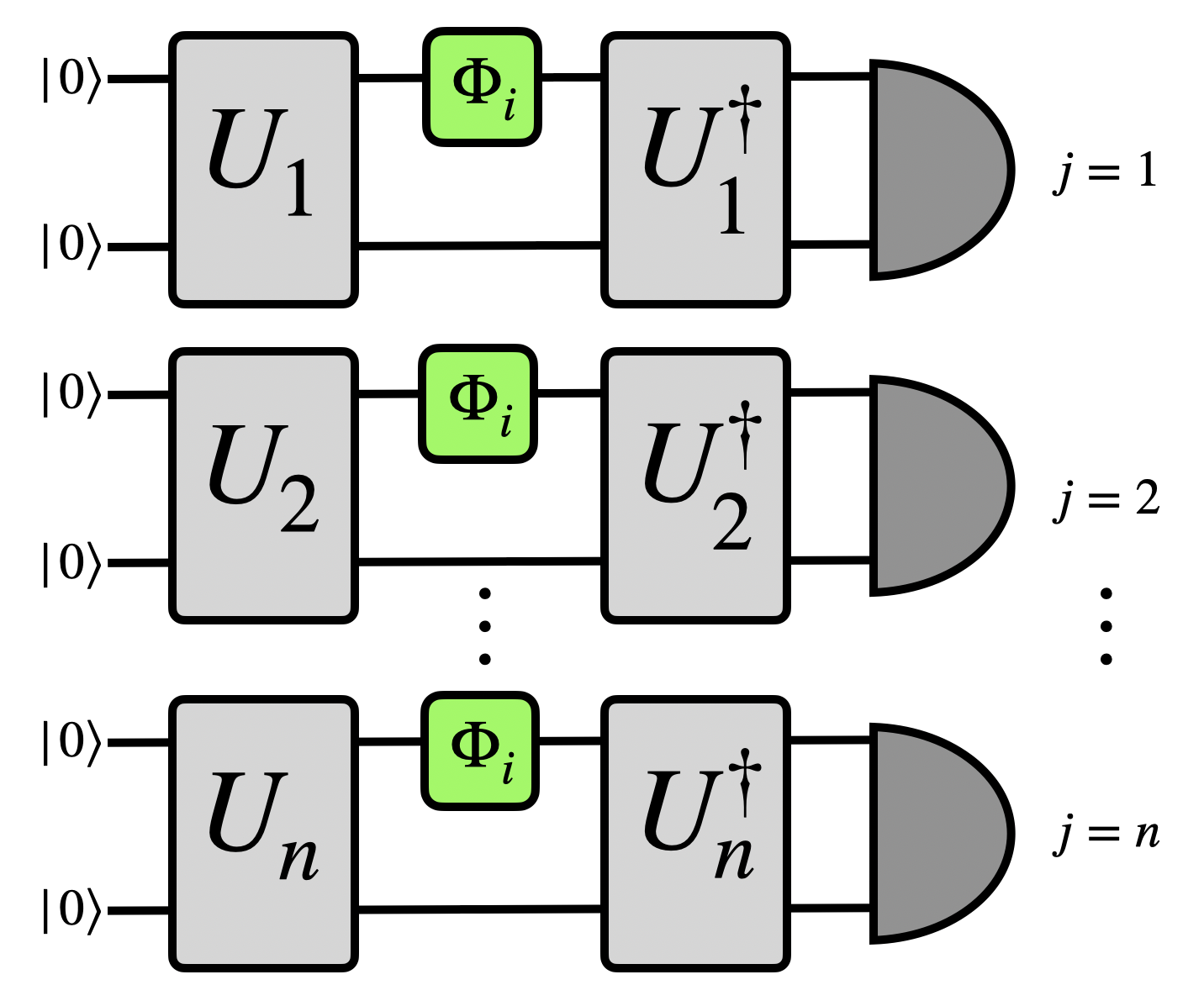}
    \caption{Schematic of QI with Haar-random probes. For each channel use $j$, a Haar-random unitary $U_j$ prepares a random signal–idler probe from the fiducial state $|00\rangle$. The signal mode is subjected to one of two competing channel hypotheses $\Phi_0$ (target absent) or $\Phi_1$ (target present), while the idler is retained. Using the classical record of the sampled unitary, the receiver applies $U_j^\dagger$ and performs the final measurement. The accumulated measurement outcomes over many independent realizations are used to distinguish the two hypotheses.
}
    \label{fig:Schematic of the protocol}
\end{figure}

The protocol considered in this work is illustrated schematically in Fig.~\ref{fig:Schematic of the protocol}. For each channel use, a Haar-random unitary prepares a signal--idler probe from a fixed fiducial state, the signal subsystem is subjected to one of the competing channel hypotheses, and the realization-dependent decoding operation is subsequently applied prior to measurement.

Let $\mathcal H_S$ and $\mathcal H_I$ denote the signal and idler Hilbert spaces, with dimensions $d_S$ and $d_I$, respectively. Fix a fiducial bipartite state $|\Omega\rangle_{SI}\in \mathcal H_S\otimes \mathcal H_I .$
For each channel use $j=1,\ldots,n$, a unitary $U_j \sim \mu_{\mathrm H}$
is sampled independently according to the Haar measure on $ {\mathsf U}(d_Sd_I).$
The corresponding probe state is $|\Psi(U_j)\rangle_{SI}
=
U_j|\Omega\rangle_{SI}.$ The classical description of the sampled unitary $U_j$ is stored in a classical register $C_j$. The associated classical–quantum state is $\Gamma(U_j)
=
|j\rangle\!\langle j|_{C_j}
\otimes
|\Psi(U_j)\rangle\!\langle\Psi(U_j)|_{SI}.$ Since the unitaries are sampled independently and identically from the Haar measure, the random variables $\{U_j\}_{j=1}^{n}$
are i.i.d. For a particular realization $\mathbf U=(U_1,\ldots,U_n),$
the corresponding $n$-copy probe state is $\Gamma^{(n)}(\mathbf U)
=
\bigotimes_{j=1}^{n}
\Gamma(U_j).$  Let $D=d_Sd_I,$
and let $|\Psi(U)\rangle
=U|\Omega\rangle,
U\sim \mu_{\mathrm H},$
where $\mu_{\mathrm H}$ denotes the normalized Haar measure on
${\mathsf U}(D)$  \cite{mele2024introduction}.
For target detection we consider the channel pair given in equations (\ref{Lloyd Channel pair_1}, \ref{Lloyd Channel pair}). After reception the receiver applies the inverse unitary $U^\dagger$.
The decoded output state under hypothesis $H_i$ is therefore
\begin{equation}
\sigma_i(U)
=
U^\dagger
(\Phi_i\otimes \id_I)
\!\left(
|\Psi(U)\rangle\!\langle\Psi(U)|
\right)
U ,
\qquad
i\in\{0,1\}.
\label{eq:sigma_i_definition}
\end{equation}
For a sequence
$
\mathbf U=(U_1,\ldots,U_n)
$
of independent Haar samples, the corresponding $n$-copy output state is
\begin{equation}
\rho_n^{(i)}(\mathbf U)
=
\bigotimes_{j=1}^{n}
\sigma_i(U_j).
\label{eq:rho_n_output}
\end{equation}
Define the single-copy Chernoff overlap
\begin{equation}
Q_s(U)
=
\operatorname{Tr}
\!\left[
\sigma_0(U)^s
\sigma_1(U)^{1-s}
\right],
\qquad
0\le s\le 1 ,
\label{eq:Qs_def}
\end{equation}
and the associated single-copy Chernoff information
\begin{equation}
X(U)
=
-\log
\Bigl(
\inf_{0\le s\le 1}
Q_s(U)
\Bigr).
\label{eq:XU_def}
\end{equation}
Since the output states are tensor products, the quantum Chernoff
distance of the realization $\mathbf U$ is

\begin{align}
\xi_n(\mathbf U)
&=
-\frac1n
\log
\inf_{0\le s\le1}
\operatorname{Tr}
\!\left[
\rho_n^{(0)}(\mathbf U)^s
\rho_n^{(1)}(\mathbf U)^{1-s}
\right]
\nonumber\\
&=
-\frac1n
\log
\inf_{0\le s\le1}
\prod_{j=1}^{n}
Q_s(U_j).
\label{eq:chernoff_realization}
\end{align}

The single-copy Chernoff exponent $X(U)$ is an i.i.d.\ random variable under Haar sampling. We therefore define
the asymptotic Haar-random Chernoff exponent as $\bar\xi
=
\mathbb E_{\mathrm H}[X(U)].$ By the strong law of large numbers \cite{mele2024introduction},
\begin{equation}
\lim_{n \to \infty}\frac1n
\sum_{j=1}^{n}
X(U_j)
\xrightarrow[]{}
\bar\xi .
\label{eq:SLLN_X}
\end{equation}
Consequently, the asymptotic discrimination performance of almost every
Haar realization is governed by the deterministic exponent
$\bar\xi$. We now show  formally that the random variable $X(U)$ concentrates sharply around
its Haar average.
Let us endow ${\mathsf U}(D)$ with the Hilbert--Schmidt metric $d_{\mathrm{HS}}(U,V)
=
\|U-V\|_2 .$ The decoded states satisfy $\sigma_i(U)
=
U^\dagger
(\Phi_i\otimes \id)
\!\left(
U\Omega U^\dagger
\right)
U ,
\Omega
=
|\Omega\rangle\!\langle\Omega|.$ Using contractivity of completely positive trace-preserving maps \cite{wilde2013quantum},
together with
\begin{align}
U\Omega U^\dagger
-
V\Omega V^\dagger
=
(U-V)\Omega U^\dagger
+
V\Omega(U^\dagger-V^\dagger),
\end{align}
one obtains
\begin{equation}
\|\sigma_i(U)-\sigma_i(V)\|_1
\le
4\|U-V\|_2 .
\label{eq:sigma_lipschitz_final}
\end{equation}
Hence the map $U\mapsto \sigma_i(U)$
is Lipschitz from
$({\mathsf U}(D),\|\cdot\|_2)$
to the trace-class operators. Because the thermal state $\rho_{\mathrm{th}}$ is full rank, every
decoded state $\sigma_i(U)$ is uniformly full rank. Therefore there
exists a constant
$\lambda_{\min}>0$
such that $\sigma_i(U)\ge \lambda_{\min} \id$
for all $U$ and for both hypotheses.

Standard Fréchet derivative bounds \cite{al2009computing} for matrix powers on the interval
$[\lambda_{\min},1]$ imply the existence of a constant $C>0$,
independent of $d$, such that
\begin{equation}
\|\sigma_i(U)^s-\sigma_i(V)^s\|_1
\le
C
\|\sigma_i(U)-\sigma_i(V)\|_1 ,
\qquad
0\le s\le1 .
\label{eq:matrix_power_lipschitz}
\end{equation}
Combining
Eqs.~\eqref{eq:sigma_lipschitz_final}
and
\eqref{eq:matrix_power_lipschitz}
gives
\begin{equation}
|Q_s(U)-Q_s(V)|
\le
L_Q
\|U-V\|_2 ,
\label{eq:Qs_lipschitz_final}
\end{equation}
with a constant $L_Q$ independent of $d$. Define $Q_*(U)
=
\inf_{0\le s\le1}Q_s(U).$
Since $\bigl|
\inf_s f_s
-
\inf_s g_s
\bigr|
\le
\sup_s |f_s-g_s|,$
Eq.~\eqref{eq:Qs_lipschitz_final} implies
\begin{equation}
|Q_*(U)-Q_*(V)|
\le
L_Q
\|U-V\|_2 .
\label{eq:Qstar_lipschitz_final}
\end{equation}

Furthermore, $Q_*(U)\ge q_{\min}>0,$ 
uniformly in $U$, because both decoded states remain uniformly
full rank. Applying the mean-value theorem to $X(U)=-\log Q_*(U)$
yields
\begin{equation}
|X(U)-X(V)|
\le
\frac{L_Q}{q_{\min}}
\,
\|U-V\|_2 .
\label{eq:X_lipschitz_final}
\end{equation}
Thus $X(U)$ is a Lipschitz function on the unitary group. 
Lévy's lemma therefore gives
\begin{equation}
\Pr
\!\left(
|X(U)-\bar\xi|
\ge \varepsilon
\right)
\le
2
\exp
\!\left(
-
\frac{c\, D^2 \varepsilon^2}
     {L_X^2}
\right),
\label{eq:levy_final}
\end{equation}
where $L_X={L_Q}/{q_{\min}}$ and $c>0$ is a universal constant. Substituting $D=d_Sd_I$ gives
\begin{equation}
\Pr
\!\left(
|X(U)-\bar\xi|
\ge \varepsilon
\right)
\le
2
\exp
\!\left(
-
\frac{
c\, d_S^2 d_I^2
}{
L_X^2
}
\varepsilon^2
\right).
\label{eq:levy_dsdi}
\end{equation}

Hence the single-copy Haar-random Chernoff information concentrates
exponentially sharply around its Haar mean. Since empirical exponents are
averages of i.i.d.\ copies of $X(U)$, standard concentration inequalities
further imply that almost every Haar realization achieves the same
asymptotic discrimination rate $\bar\xi$.
\paragraph{Unitary channel discrimination.}

A particularly transparent situation arises when the competing channels
are unitary, $\Phi_0(\rho)
=
V_0 \rho V_0^\dagger,
\Phi_1(\rho)
=
V_1 \rho V_1^\dagger .$
Define the relative unitary $W:=V_0^\dagger V_1 .$ For a Haar-random encoding unitary $U\in{\mathsf U}(D),$
the probe state is $|\Psi(U)\rangle
=
U|\Omega\rangle .$ After channel application and decoding, the corresponding pure output
states are $|\psi_0(U)\rangle
=
U^\dagger
(V_0\otimes \mathbb{I}_I)
U
|\Omega\rangle ,
|\psi_1(U)\rangle
=
U^\dagger
(V_1\otimes \mathbb{I}_I)
U
|\Omega\rangle .$ Their overlap is
\begin{equation}
\langle\psi_0(U)|\psi_1(U)\rangle
=
\langle\Omega|
U^\dagger
(W\otimes \mathbb{I}_I)
U
|\Omega\rangle .
\label{eq:haar_overlap}
\end{equation}
The decoded output states introduced in
Eq.~\eqref{eq:sigma_i_definition}
are therefore rank-one projectors,
\begin{equation}
\sigma_i(U)
=
|\psi_i(U)\rangle
\langle\psi_i(U)| ,
\qquad i\in\{0,1\}.
\end{equation}

For pure states, $\operatorname{Tr}
\!\left[
\sigma_0(U)^s
\sigma_1(U)^{1-s}
\right]
=
\bigl|
\langle\psi_0(U)|\psi_1(U)\rangle
\bigr|^2 ,$
independently of $s$ \cite{Audenaert07}. Hence the single-copy Chernoff overlap becomes
\begin{equation}
Q_s(U)
=
|Z(U)|^2 ,
\label{eq:unitary_Qs}
\end{equation}
where $Z(U)
=
\langle\Omega|
U^\dagger
(W\otimes \mathbb{I}_I)
U
|\Omega\rangle .$ 
Consequently,
\begin{equation}
X(U)
=
-\log |Z(U)|^2 .
\label{eq:unitary_X}
\end{equation}
The statistics of the Haar-random Chernoff information are therefore
completely determined by the random quadratic form $Z(U)$.

Its second moment may be evaluated using standard Haar-integration
formulas and Weingarten calculus
\cite{mele2024introduction},
yielding
\begin{equation}
\mathbb E_{\mathrm H}
\!\left[
|Z(U)|^2
\right]
=
\frac{
\bigl|
\operatorname{Tr}(W\otimes \mathbb{I}_I)
\bigr|^2
+D
}
{D(D+1)} ,
\label{eq:haar_second_moment}
\end{equation}
where $D=d_Sd_I$.

Since $\operatorname{Tr}(W\otimes \mathbb{I}_I)
=
d_I\,\operatorname{Tr}(W),$ 
Eq.~\eqref{eq:haar_second_moment} may also be written as
\begin{equation}
\mathbb E_{\mathrm H}
\!\left[
|Z(U)|^2
\right]
=
\frac{
d_I^2 |\operatorname{Tr}(W)|^2
+D
}
{D(D+1)} .
\end{equation}

For fixed idler dimension and large total dimension
$D=d_Sd_I$, $\mathbb E_{\mathrm H}
\!\left[
|Z(U)|^2
\right]
=1/D
+
O(D^{-2}),$ 
provided
$
|\operatorname{Tr}(W)|=O(1).$  Here
$O(D^{-2})$ represents corrections whose magnitude is bounded by
$C D^{-2}$ for some dimension-independent constant $C$ in the
large-$D$ limit. Thus, the typical overlap obeys $|Z(U)|
=
O
\!\left(
D^{-1/2}
\right),$ 
and therefore
\begin{equation}
X(U)
=
-\log |Z(U)|^2
=
\log D
+
O(1).
\end{equation}

Since $X(U)$ concentrates around its Haar mean according to
Eq.~\eqref{eq:levy_dsdi}, the Haar-random Chernoff exponent satisfies
\begin{equation}
\bar{\xi}
=
\mathbb E_{\mathrm H}[X(U)]
=
\log D
+
O(1),
\qquad
D\rightarrow\infty .
\label{eq:haar_scaling}
\end{equation}

The leading term is universal and depends only on the accessible
Hilbert-space dimension. Dependence on the specific channel pair enters
through the subleading correction governed by the spectral form factor $K_1(W)
=
|\operatorname{Tr}(W)|^2 .$ Thus, in the unitary channel-discrimination setting, Haar-random quantum
illumination acquires a direct connection to spectral form factor \cite{SFF} and
quantum chaos through the relative unitary $W=V_0^\dagger V_1$.  {The spectral form factor of the relative unitary acquires an operational meaning: it quantifies how readily two many-body quantum evolutions can be distinguished using Haar-random probes.
}

{\paragraph{General completely-positive-trace-preserving (CPTP) channel discrimination.}

For two CPTP maps $\Phi_0$ and $\Phi_1$, let
\begin{equation}
\sigma_i(U)
=
U^\dagger
(\Phi_i\otimes\id_I)
\!\left[
|\Psi(U)\rangle\!\langle\Psi(U)|
\right]U,
i\in\{0,1\},
\end{equation}
denote the decoded output states generated from the Haar-random probe
$|\Psi(U)\rangle=U|\Omega\rangle$. The single-copy
Chernoff overlap is defined as before. 
The Haar-random Chernoff exponent is therefore $\bar{\xi}
=
\mathbb E_{\mathrm H}[X(U)].$ Unlike the unitary case, there is generally no reduction to a single
relative operator. Instead, the Haar average is governed by the Choi
states $J(\Phi_0)$ and $J(\Phi_1)$. Using the second-moment Haar
identity \cite{mele2024introduction}, $\mathbb E_{\mathrm H}
\!\left[
|\Psi(U)\rangle\!\langle\Psi(U)|^{\otimes2}
\right]
=
({\id +F})/\{{D(D+1)}\},$ (where \(F\) denotes the swap (flip) operator on \(\mathcal H\otimes\mathcal H\), defined by \(F(|\phi\rangle\otimes|\psi\rangle)=|\psi\rangle\otimes|\phi\rangle\)) the leading Haar-averaged distinguishability is controlled by quadratic
invariants of the Choi difference
$\Delta J
=
J(\Phi_0)-J(\Phi_1).$
Consequently, in the large-$D$ limit,
$\bar{\xi}
=
\log D
+
O(1),
D\rightarrow\infty,$
where the subleading correction depends on channel-specific invariants
constructed from $\Delta J$. Thus, for general CPTP maps, Haar-random
QI probes the geometry of channel distinguishability
encoded in the corresponding Choi state representations, playing the role
analogous to the spectral form factor $|\operatorname{Tr}(W)|^2$ in the
unitary setting. This brings up the role of fluctuations of spectral form factor in the hypothesis testing task. }

\subsection{Symmetry-Constrained Haar-random Probe States}
\label{sec:symmetry_haar_chernoff}

We now generalize the Haar-random illumination protocol by
restricting the probe ensemble to a fixed symmetry sector of
the signal--idler Hilbert space.  Rather than sampling Haar-random
probe states from the full Hilbert space, the random unitaries
are constrained to act within a chosen irreducible
representation sector. The resulting discrimination exponent
may therefore be interpreted as a symmetry-resolved version
of the Haar-random Chernoff exponent introduced above.

Let a compact group $G$ act on the signal--idler Hilbert space
$\mathcal H$ through a unitary representation
$g\mapsto U_g$.
The Hilbert space decomposes as $\mathcal H
=
\bigoplus_{\lambda}
\mathcal H_{\lambda},$
where $\mathcal H_\lambda$ denotes an irreducible
representation sector of dimension $d_\lambda
=
\dim(\mathcal H_\lambda).$ Fix a fiducial state $|\Omega_\lambda\rangle
\in
\mathcal H_\lambda .$ For each channel use, a Haar-random unitary $U_\lambda
\in
{\mathsf  U}(d_\lambda)$ is sampled according to the normalized Haar measure
$\mu_\lambda$ \cite{mele2024introduction} on the sector
${\mathsf  U}(d_\lambda)$.
The corresponding probe state is $|\Psi_\lambda(U_\lambda)\rangle
=
U_\lambda |\Omega_\lambda\rangle .$ As in the unrestricted protocol, the realization
$U_\lambda$ is stored classically and used for decoding
after the channel action. The decoded single-copy output state
under hypothesis $H_i$ is
\begin{equation}
\sigma_i^{(\lambda)}(U_\lambda)
=
U_\lambda^\dagger
\left[\Phi_i\otimes\id
\!\left(
|\Psi_\lambda(U_\lambda)\rangle
\langle
\Psi_\lambda(U_\lambda)|
\right)
\right]
U_\lambda,~~
i\in\{0,1\}.
\label{eq:symmetry_sigma}
\end{equation}

Exactly as in the unrestricted case, the asymptotic
discrimination problem is determined by the Haar-random pair
$
(\sigma_0^{(\lambda)}(U_\lambda),
 \sigma_1^{(\lambda)}(U_\lambda))
$. As before,  define the symmetry-resolved Chernoff overlap
\begin{equation}
Q_s^{(\lambda)}(U_\lambda)
=
\operatorname{Tr}
\!\left[
\bigl(
\sigma_0^{(\lambda)}(U_\lambda)
\bigr)^s
\bigl(
\sigma_1^{(\lambda)}(U_\lambda)
\bigr)^{1-s}
\right],
0\le s\le1 ,
\label{eq:symmetry_Qs}
\end{equation}
and the corresponding single-copy Chernoff information
\begin{equation}
X_\lambda(U_\lambda)
=
-\log
\left(
\inf_{0\le s\le1}
Q_s^{(\lambda)}(U_\lambda)
\right).
\label{eq:symmetry_X}
\end{equation}
The asymptotic symmetry-resolved Chernoff exponent is then
\begin{equation}
\bar{\xi}_\lambda
=
\mathbb E_{\mu_\lambda}
\!\left[
X_\lambda(U_\lambda)
\right],
\label{eq:symmetry_average_exponent}
\end{equation}
where the expectation is taken with respect to the Haar measure
on ${\mathsf  U}(d_\lambda)$. For $n$ independent realizations
$
(U_{\lambda,1},\ldots,U_{\lambda,n})
$,
the empirical Chernoff exponent converges almost surely to
$\bar{\xi}_\lambda$ by the strong law of large numbers,
exactly as in the unrestricted Haar-random protocol.

If both channels are $G$-covariant, $\Phi_i\circ\mathcal U_g
=
\mathcal U_g\circ\Phi_i,
\forall g\in G,$
where
$
\mathcal U_g(\rho)
=
U_g \rho U_g^\dagger,
$
then Schur's lemma implies that distinct irreducible sectors
remain dynamically decoupled. Consequently, the channel
discrimination problem decomposes independently within each
symmetry sector. If the preparation sector may be chosen freely, the optimal
asymptotic discrimination rate is therefore
\begin{equation}
\bar{\xi}_{\mathrm{opt}}
=
\max_{\lambda}
\bar{\xi}_\lambda .
\label{eq:optimal_sector_rate}
\end{equation}

Thus symmetry provides a natural decomposition of the
Haar-random discrimination problem into a collection of
sector-resolved Chernoff exponents, each associated with a
particular irreducible representation of the underlying
symmetry group.

 If the channels are not $G$-covariant, the dynamics generally mixes distinct irreducible symmetry sectors. Nevertheless, the symmetry averaging inherent in the twirling procedure removes all components that are not invariant under the adjoint action of $G$. For a channel $\mathcal{E}$, the corresponding symmetry-twirled channel is defined as

\begin{equation}
\mathcal{E}^{(G)}
=\int_G dg
\mathcal{U}_g^\dagger \circ \mathcal{E} \circ \mathcal{U}_g ,
\qquad
\mathcal{U}_g(\rho)=U_g \rho U_g^\dagger,
\end{equation}
where $dg$ denotes the normalized Haar measure on $G$. The twirling operation projects $\mathcal{E}$ onto the subspace of channels that are invariant under the group action and therefore retains only its $G$-covariant component. Consequently, only the $G$-covariant components of the channels contribute to the Haar-averaged Chernoff exponent, while symmetry-breaking components are eliminated by the group average. {Our analysis is exemplified by permutational invariant states, enjoyed by a variety of quantum condensed matter and quantum optics models. While these models serve as sources, the channel may or maynot enjoy the symmetry, both cases covered by our analysis.}

\subsubsection{Unitary Channels within a Symmetry Sector}
For unitary channels $\Phi_j(\rho)=V_j\rho V_j^\dagger$ $(j=0,1)$, define the relative unitary $W=V_0^\dagger V_1$ and let the Haar-random probe in the symmetry sector $\mathcal H_\lambda$ be $|\Psi_\lambda(U_\lambda)\rangle=U_\lambda|\Omega_\lambda\rangle$, with $U_\lambda\in{\mathsf  U}(d_\lambda)$. After channel application and decoding, the output states are $|\psi_j^{(\lambda)}(U_\lambda)\rangle=U_\lambda^\dagger (V_j\otimes \mathbb{I}_I)U_\lambda|\Omega_\lambda\rangle$ for $j=0,1$.
As before, their overlap is
\begin{equation}
\langle
\psi_0^{(\lambda)}(U_\lambda)
|
\psi_1^{(\lambda)}(U_\lambda)
\rangle
=
\langle
\Omega_\lambda
|
U_\lambda^\dagger
(W\otimes \mathbb{I}_I)
U_\lambda
|
\Omega_\lambda
\rangle .
\label{eq:symmetry_overlap}
\end{equation}
The decoded states
$
\sigma_i^{(\lambda)}(U_\lambda)
$
are rank-one projectors,
\begin{equation}
\sigma_i^{(\lambda)}(U_\lambda)
=
|\psi_i^{(\lambda)}(U_\lambda)\rangle
\langle
\psi_i^{(\lambda)}(U_\lambda)
| .
\end{equation}
Following previous arguments, 
\begin{equation}
X_\lambda(U_\lambda)
=
-\log
|Z_\lambda(U_\lambda)|^2 .
\label{eq:Xlambda_unitary}
\end{equation}
Thus the symmetry-resolved discrimination problem depends only
on the restriction of the relative unitary $W$ to the chosen
symmetry sector.

\subsubsection{Large-Sector Asymptotics}

Let $W_\lambda
=
\Pi_\lambda
(W\otimes \mathbb{I}_I)
\Pi_\lambda ,$
denote the restriction of the relative unitary to the sector
$\mathcal H_\lambda$, where $\Pi_\lambda$ is the projector
onto $\mathcal H_\lambda$. Standard Haar-integration formulas based on Weingarten calculus
\cite{mele2024introduction} give
\begin{equation}
\mathbb E_{\mu_\lambda}
\!\left[
|Z_\lambda(U_\lambda)|^2
\right]
=
\frac{
|\operatorname{Tr}(W_\lambda)|^2
+
d_\lambda
}
{
d_\lambda(d_\lambda+1)
}.
\label{eq:sector_second_moment}
\end{equation}

For generic relative unitaries satisfying $|\operatorname{Tr}(W_\lambda)|^2
=
O(1),$ one obtains $\mathbb E_{\mu_\lambda}
\!\left[
|Z_\lambda(U_\lambda)|^2
\right]
={1}/{d_\lambda}
+
O(d_\lambda^{-2}).$
Hence the typical overlap scales as $|Z_\lambda(U_\lambda)|
=
O_{\mathrm{typ}}
\!\left(
d_\lambda^{-1/2}
\right),$
which implies 
 $X_\lambda(U_\lambda)
=
-\log
|Z_\lambda(U_\lambda)|^2
=
\log d_\lambda
+
O(1).$
Since $X_\lambda(U_\lambda)$ is a Lipschitz function on
${\mathsf  U}(d_\lambda)$, the same concentration-of-measure
arguments used in the unrestricted Haar-random setting imply that
$X_\lambda(U_\lambda)$ concentrates exponentially around its
sector average $\bar{\xi}_\lambda
=
\mathbb E_{\mu_\lambda}
\!\left[
X_\lambda(U_\lambda)
\right].$ Consequently,
\begin{equation}
\bar{\xi}_\lambda
=
\log d_\lambda
+
O(1),
\qquad
d_\lambda\rightarrow\infty .
\label{eq:sector_large_d}
\end{equation}
Once more, the leading asymptotic contribution depends only on the
dimension of the accessible symmetry sector.
The subleading corrections are controlled by the
symmetry-resolved spectral form factor $K_\lambda(W)
=
|\operatorname{Tr}(W_\lambda)|^2 .$ Thus symmetry-resolved channel discrimination provides an
operational interpretation of symmetry-resolved spectral form
factors and their associated notions of quantum chaos.
\subsubsection{Comparison Between Symmetry-Constrained and Unconstrained Probes}

For unrestricted Haar-random probes acting on the full Hilbert space
$\mathcal H$ of dimension
$D=d_Sd_I$,
the asymptotic Chernoff exponent satisfies $\bar{\xi}
=
\log D
+
O(1).$ When probes are restricted to an irreducible symmetry sector
$\mathcal H_\lambda$ of dimension $d_\lambda$, the corresponding
symmetry-resolved exponent is $\bar{\xi}_\lambda
=
\log d_\lambda
+
O(1).$ Hence
\begin{equation}
\bar{\xi}_\lambda-\bar{\xi}
=
\log\!\left(
\frac{d_\lambda}{D}
\right)
+
O(1),
\label{eq:rate_difference}
\end{equation}
showing that the leading effect of the symmetry constraint is simply
the reduction of the accessible Hilbert-space dimension.

The subleading corrections are governed by the symmetry-resolved
spectral form factor $K_\lambda(W)
=
\left|
\operatorname{Tr}(W_\lambda)
\right|^2,$
where $W_\lambda
=
\Pi_\lambda
(W\otimes \mathbb{I}_I)
\Pi_\lambda$
is the restriction of the relative unitary to the symmetry sector
$\mathcal H_\lambda$. For chaotic dynamics one expects random-matrix
statistics, implying
$K(W)=O(D)$ and
$K_\lambda(W)=O(d_\lambda)$.
These contributions modify only the $O(1)$ terms and do not affect
the leading logarithmic scaling.

A useful dimensionless measure of the symmetry-induced reduction is $\eta_\lambda
=
{\bar{\xi}_\lambda}/{\bar{\xi}}
={\log d_\lambda}/{\log D}
+
O(1),$ 
which quantifies the fraction of the unrestricted discrimination rate
retained within the symmetry sector. For many-body systems, the dimensions of common symmetry sectors scale
as $d_N = \binom{L}{N},
d_M = \binom{L}{L/2+M},\\
d_k \sim {2^L}/{L},$ 
where $N$ denotes the conserved particle number, $M$ the total
magnetization, and $k$ the crystal momentum quantum number \cite{larocca2022diagnosing}. The
corresponding asymptotic exponents are therefore
\begin{align}
\bar{\xi}_{N}
&\sim
\log
\binom{L}{N},
\\
\bar{\xi}_{M}
&\sim
\log
\binom{L}{L/2+M},
\\
\bar{\xi}_{k}
&\sim
L\log 2
-
\log L ,
\end{align}
up to subleading corrections determined by
$K_\lambda(W)$.

At half filling,
$N=L/2$,
Stirling's approximation yields
\begin{equation}
\log
\binom{L}{L/2}
=
L\log 2
-
\frac12\log L
+
O(1),
\end{equation}
so that $\bar{\xi}_{N}
=
\bar{\xi}
-\log L/2
+
O(1).$
Similarly, $\bar{\xi}_{M}
=
L\log 2
-\log L/2
+
O(1),
\bar{\xi}_{k}
=
L\log 2
-
\log L
+
O(1).$ 
Thus, although symmetry restrictions reduce the accessible Hilbert
space, the associated loss in discrimination power is only
subextensive in system size. The general conclusion is
\begin{equation}
\bar{\xi}_{\mathrm{sym}}
\sim
\log d_{\mathrm{acc}},
\end{equation}
where $d_{\mathrm{acc}}$ denotes the dimension of the accessible
symmetry sector. Symmetry therefore preserves the logarithmic form of
the asymptotic Chernoff exponent while replacing the full Hilbert-space
dimension by the effective dimension available within the chosen
sector.

{The above analysis extends directly to quantum channels that preserve the underlying symmetry. In particular, one may consider completely positive trace-preserving maps whose Kraus operators act within a fixed symmetry sector $\mathcal H_\lambda$, so that the channel dynamics remains block diagonal and can be analyzed independently in each sector.}

{Having established how symmetry constraints modify the asymptotic Chernoff exponent through the effective dimension of the accessible Hilbert space, we now turn to QI. As discussed above, since QI is a channel-discrimination problem, the preceding analysis suggests that its performance should likewise be governed by the dimension of the symmetry sector from which probe states are drawn. We therefore investigate whether the quantum advantage of illumination survives when the probes are restricted to symmetry-constrained Haar-random ensembles.

\section{Quantum Illumination with Symmetry-Constrained Haar-random States}
\label{sec:QI_symmetry_constrained}
{ Let $Q_S$ and $Q_I$ be local conserved charges satisfying $Q_S|q\rangle_S=q|q\rangle_S,
Q_I|q'\rangle_I=q'|q'\rangle_I,$
and define the global conserved quantity $Q=Q_S-Q_I.$
The Hilbert space decomposes into charge sectors, $\mathcal H=\bigoplus_{\lambda}\mathcal H_\lambda.$
We focus on the zero-charge sector, 
$\mathcal H_0\equiv \mathcal H_{Q=0},$ whose dimension is $d_0=\dim(\mathcal H_0).$
A convenient basis is $\mathcal H_0
=
\operatorname{span}
\{
|k\rangle_S|k\rangle_I
:
k=1,\ldots,d_0
\}.$} Choose the fiducial state $|\Omega_0\rangle
=
\sum_{k=1}^{d_0}
|k\rangle_S|k\rangle_I/{\sqrt{d_0}}.$ A typical  Haar-random probe state within the symmetry sector is then $|\Psi_0(U_0)\rangle
=
U_0
|\Omega_0\rangle ,
U_0
\in
 {\mathsf U}(d_0),$
with $U_0$ distributed according to the Haar measure
on $ {\mathsf U}(d_0)$ and $|\Omega_0\rangle$ is a fiducial state. Equivalently,  Haar random pure states can be written as $|\Psi_0\rangle
\approx
\sum_{k=1}^{d_0}
e^{i\theta_k}
|k\rangle_S|k\rangle_I/ {\sqrt{d_0}},$ where the phases $\{\theta_k\}$ are independent and uniformly
distributed on $[0,2\pi)$ \cite{mele2024introduction}. These states satisfy (a) exact charge conservation,      $Q|\Psi_0\rangle=0,$ (b) maximal Schmidt rank $d_0$, and (c) maximally mixed reduced states,  $\rho_S=\rho_I={\mathbb I}/{d_0}.$ This approximation captures the
delocalized support and random phases of a typical Haar state within
$\mathcal H_0$, while neglecting the Porter--Thomas fluctuations of the
amplitudes, which contribute only subleading $O(1)$ corrections to the
large-$d_0$ scaling considered below.

We employ the same quantum-illumination model introduced in
Sec.~\ref{subsec:quantum_illumination_channel_discrimination}. Under hypothesis $H_0$ (target absent), the signal is replaced
by thermal noise, $\rho_{SI}^{(0)}
=
\rho_{\mathrm{th}}
\otimes{\mathbb I_I}/{d_0}.$ Under hypothesis $H_1$ (target present), $\rho_{SI}^{(1)}
=
(1-\eta)\rho_{SI}^{(0)}
+
\eta
|\Psi_0\rangle
\langle\Psi_0|.$ We consider the binary
measurement $\Pi_{\mathrm{yes}}
=
|\Psi_0\rangle
\langle\Psi_0|,
\Pi_{\mathrm{no}}
=
\mathbb I
-
\Pi_{\mathrm{yes}} .$ 
{ To leading order in $b$ and $\eta$, this measurement is optimal because $\Pi_{\mathrm{yes}}=|\Psi_0\rangle\langle\Psi_0|$ is precisely the projector onto the positive part of $\rho_{SI}^{(1)}-\rho_{SI}^{(0)}$, i.e., the Helstrom measurement.} Under $H_0$,
\begin{align}
p_0
&=
\operatorname{Tr}
\!\left[
\Pi_{\mathrm{yes}}
\rho_{SI}^{(0)}
\right]=
\operatorname{Tr}
\!\left[
|\Psi_0\rangle
\langle\Psi_0|
\left(
\rho_{\mathrm{th}}
\otimes
\frac{\mathbb I_I}{d_0}
\right)
\right].
\end{align}

Restricting to the single-photon sector yields $p_0
={b}/{d_0}.$ Under $H_1$,

\begin{align}
p_1
&=
(1-\eta)p_0+\eta
=
\eta
+
(1-\eta)
\frac{b}{d_0}.
\label{eq:p1_sector}
\end{align}

The quantum hypothesis testing problem  therefore reduces to a
classical binary hypothesis test between Bernoulli
distributions $P_0
=
(p_0,1-p_0),
P_1
=
(p_1,1-p_1).$

We now calculate the Chernoff bound. 
For $n$ independent repetitions, the optimal Bayesian error
probability satisfies

\begin{align}
P_{\mathrm{err}}^{(n)}
=
\exp
\!\left[
-n\bar{\xi}_0
+
O(n)
\right],
\end{align}

where

\begin{equation}
\bar{\xi}_0
=
-\log
\left[
\min_{0\le s\le1}
\Big(
p_0^s p_1^{1-s}
+
(1-p_0)^s(1-p_1)^{1-s}
\Big)
\right].
\label{eq:Chernoff_lambda}
\end{equation}

Define $\varepsilon_0
={b}/{d_0}.$ Then $p_0=\varepsilon_0,
p_1
=
\eta
+
(1-\eta)\varepsilon_0 .$

\paragraph*{Good regime:
$\eta\gg\varepsilon_0$}

When the reflectivity exceeds the effective noise level,

\begin{equation}
\bar{\xi}_0
=
\eta
+
O\!\left(
\frac{b}{d_0}
\right).
\label{eq:good_regime_lambda}
\end{equation}

The leading exponent coincides with the exponent for the standard
quantum-illumination protocol \cite{lloyd2008enhanced}.

\paragraph*{Bad regime:
$\eta\ll\varepsilon_0$}

Expanding the Chernoff information for nearby Bernoulli
distributions gives

\begin{equation}
\bar{\xi}_0
=
\frac{\eta^2 d_0}
     {8b}
+
O(\eta^3).
\label{eq:bad_regime_lambda}
\end{equation}

Thus the symmetry-sector dimension enhances the discrimination
exponent linearly. Combining the two regimes yields

\begin{equation}
\bar{\xi}_0
\simeq
\begin{cases}
\eta,
&
\eta\gg b/d_0,
\\
\dfrac{\eta^2 d_0}{8b},
&
\eta\ll b/d_0.
\end{cases}
\label{eq:QI_sector_exponent}
\end{equation}

Equation~\eqref{eq:QI_sector_exponent} is identical to the
standard quantum-illumination exponent \cite{lloyd2008enhanced} after replacing
the full Schmidt rank by the symmetry-sector dimension
$d_0$. The operational resource governing the asymptotic error exponent
is therefore not the total Hilbert-space dimension but the
dimension of the symmetry sector supporting the probe state.

The above expressions coincide exactly with those obtained for the maximally entangled probe. In particular, the effective noise probability is suppressed from $b$ to $b/d_0$. Before proceeding, we emphasize that the present analysis is carried out within the simplified quantum-illumination model introduced in \cite{lloyd2008enhanced}. In this model, the presence or absence of the target is modeled by a weakly reflecting beam splitter with reflectivity $\eta$, which mixes the transmitted signal with a thermal environment. In a more realistic description, the scattering operation associated with the unknown object need not preserve the symmetry sector in which the probe is prepared and may therefore couple different charge sectors. Consequently, the returned state can exhibit symmetry-breaking components even when the incident probe lies entirely within a fixed symmetry sector. Since the final detection is made at the receiver, however, one may equivalently absorb such effects into the detection stage and project onto the desired symmetry sector. The results derived above should therefore be interpreted as characterizing the symmetry-resolved component of the quantum-illumination protocol within the effective weak, symmetry non-preserving beam-splitter model.
\section{Charge-Conserving Haar-Random Hamiltonian Models}
\label{sec:Discussion}
A broad class of physically relevant generators is formed by hopping and interaction Hamiltonians containing equal numbers of creation and annihilation operators. For bosonic lattice systems, a representative example is the Bose--Hubbard Hamiltonian \cite{batrouni1995supersolids,bosehubbard_expt},
\begin{align}
H=-\sum_{\langle i,j\rangle} t_{ij}\left(a_i^\dagger a_j+\mathrm{H.c.}\right)+\sum_{i,j}V_{ij}n_i n_j ,
\end{align}
where $a_i^\dagger$ ($a_i$) creates (annihilates) a boson on site $i$, $n_i=a_i^\dagger a_i$ is the local number operator, $t_{ij}$ is the hopping amplitude between sites $i$ and $j$, $V_{ij}$ is the density--density interaction strength, and $\mathrm{H.c.}$ denotes the Hermitian conjugate. For fermionic lattice systems, a corresponding example is the Fermi--Hubbard Hamiltonian \cite{cade2020strategies, wang2022experimental, xu2023frustration},
\begin{align}
H=-\sum_{\langle i,j\rangle,\sigma}t_{ij}\left(c_{i\sigma}^\dagger c_{j\sigma}+\mathrm{H.c.}\right)+U\sum_i n_{i\uparrow}n_{i\downarrow},
\end{align}
where $c_{i\sigma}^\dagger$ ($c_{i\sigma}$) creates (annihilates) a fermion with spin $\sigma$ on site $i$, $n_{i\sigma}=c_{i\sigma}^\dagger c_{i\sigma}$, and $U$ is the on-site interaction strength. Both Hamiltonians satisfy $ [H,\hat N]=0 $, and therefore generate number-conserving unitary evolutions belonging to the symmetry-restricted compact unitary group associated with particle-number conservation. An experimentally accessible photonic realization is provided by cavity-QED systems of ultracold bosons coupled to a single optical cavity mode \cite{halati2022breaking,nonlinearcavity_expt}. Because these models are non-integrable, they generate random unitaries.  The resulting strong symmetry sectors generate symmetry-constrained dynamics, making such platforms a natural setting for experimentally probing the hypothesis-testing and channel-discrimination protocols considered in this work. 

It is instructive to compare the present results with the randomness-assisted illumination protocol employing  coherent states~\cite{brougham2023using}. There, coherent states with randomly chosen intensities are sampled from a classical distribution whose ensemble average reproduces a thermal state. By retaining classical information about the realization and incorporating it into the detection procedure, it was shown that several features commonly associated with direct-measurement QI can be reproduced without employing entanglement or signal--idler correlations. This observation demonstrates that randomness, when accompanied by suitable side information, can itself become a useful resource for discrimination tasks.

The framework developed here reveals a complementary mechanism. Rather than exploiting classical randomness in the preparation of coherent states, the present protocol employs Haar-random probe states drawn from symmetry-resolved sectors of the Hilbert space. The relevant side information is the classical record of the sampled Haar unitary, which permits decoding into a common reference frame prior to measurement. Consequently, the resulting discrimination advantage is governed not by fluctuations in probe intensity, but by the geometry and dimension of the accessible correlation subspace.
 From this viewpoint, both protocols suggest a broader interpretation of QI. The essential ingredient is not necessarily a particular engineered entangled state, but the existence of a sufficiently large ensemble of distinguishable configurations together with auxiliary information that allows those configurations to be resolved at the receiver. In the present setting, this structure is quantified by the symmetry-sector dimension $d_0$, which directly controls the asymptotic Chernoff exponent.
\section{Conclusion and outlook}
\label{sec:Conclusion and outlook}
In this work, we developed a symmetry-resolved framework for quantum channel discrimination using Haar-random probe states. By formulating Haar-random probe ensembles within a classical--quantum setting that retains the realization-dependent decoding information, we obtained an exact i.i.d.\ discrimination protocol whose asymptotic performance is governed by a quantum-Chernoff exponent. We showed that the resulting Haar-random Chernoff information concentrates exponentially around its Haar average, implying that almost every realization achieves the same asymptotic discrimination rate.

For unrestricted Haar-random probes acting on a Hilbert space of dimension $d$, we found that the asymptotic Chernoff exponent scales as
$\bar{\xi}\sim \log d,$
up to channel-dependent subleading corrections. Restricting the probe ensemble to a symmetry sector $\mathcal H_\lambda$ of dimension $d_\lambda$ yields the symmetry-resolved exponent $\bar{\xi}_\lambda\sim \log d_\lambda.$ 
Thus the leading asymptotic discrimination rate is determined not by the full Hilbert-space dimension but by the dimension of the accessible symmetry sector. For unitary channel discrimination, the subleading corrections are governed by the symmetry-resolved spectral form factor $K_\lambda(W)=|\mathrm{Tr}(W_\lambda)|^2,$
providing an operational interpretation of symmetry-resolved spectral statistics and establishing a direct connection between channel discrimination, spectral form factors, and quantum chaos. The same perspective extends naturally to symmetry-preserving CPTP channels, where the relevant channel-dependent corrections are encoded in symmetry-resolved invariants of the corresponding Choi states. These results reveal that symmetry-resolved channel discrimination provides a natural operational probe of the geometry of quantum channels within conserved-charge sectors. 
As an application of this general framework, we investigated QI with symmetry-constrained Haar-random states. We showed that Haar-random probes drawn from a fixed symmetry sector reproduce the characteristic noise suppression of conventional QI, with the effective thermal-noise contribution reduced from $b$ to $b/d_\lambda$. Consequently, the quantum-illumination Chernoff exponent is determined by the dimension of the accessible symmetry sector and coincides with the standard result after replacing the Schmidt rank by $d_\lambda$. This demonstrates that the operational resource underlying QI is not a specially engineered probe state but rather the existence of a large symmetry-resolved correlation subspace supporting distinguishable channel responses.

Our results therefore shift the emphasis from state engineering to subspace engineering. The achievable discrimination rate is governed primarily by the geometry and dimension of the symmetry sector accessible to the probe ensemble, while typical states within that sector already attain near-optimal performance. This perspective places QI within a broader framework connecting symmetry-resolved typicality, quantum channel discrimination, many-body quantum chaos, and spectral statistics. Several directions remain open. An important question is whether one can identify an operationally optimal symmetry sector directly from the structure of the competing channels, for example through symmetry-resolved Choi states or channel distinguishability measures. It would also be interesting to extend the present analysis to more general symmetry-preserving noise models, adaptive discrimination protocols, and experimentally accessible many-body systems exhibiting conserved quantities and chaotic dynamics. Such investigations may further clarify the role of symmetry-resolved Hilbert-space geometry as a resource for quantum-enhanced sensing and channel discrimination.

\section*{Acknowledgement}

SV acknowledges funding under
the Government of India’s National Quantum Mission grants numbered DST/QTC/NQM/QC/2024/1 and
DST/FFT/NQM/QSM/2024/3. SV also acknowledges
useful discussions at the International Centre for Theoretical Sciences (ICTS) during the programs with codes
ICTS/qm1002025/01 and ICTS/qt2025/01 and discussions with S. Guha.
While preparing this manuscript, we became aware of the recent preprint arXiv:2605.31168 \cite{marcinkowska2026asymptotic} where the authors note a suppression of type-II error using Haar unitaries.

%

\end{document}